# Is it possible to "prescribe" the diffusion for a geminate pair in a force field? [¶]

Ilya A. Shkrob

*[a] Chemistry Division, Argonne National Laboratory, Argonne, IL 60439*

## Abstract

Given the difficulty of obtaining compact analytical solutions for diffusion of interacting geminate pairs (such as electron-hole pairs generated by ionization of liquid) it is common, following the original method of Mozumder, to "prescribe" this diffusion. With this approach, the survival probability of the pair is represented as a product of the survival probability for a freely diffusing pair and a suitably defined weighting function. This approach gives the correct limiting survival probability for a pair in the Coulomb field. The same approach was used for simulation of reaction dynamics in radiolytic spurs ("independent reaction times" approach of Pimblott) and solve other vexing diffusion problems that do not have analytical solution. A reasonable question is, can the same method be used for any other interaction potential than Coulomb? Here we demonstrate that such a prescription is generally impossible. The correct result given by the prescribed diffusion approach for the Coulomb potential is, actually, purely accidental. The method is inherently contradictory and it should be used with caution.

Dissociation, ionization, and electron photodetachemnt frequently yield geminate pairs in which the partners recombine and interact with each other. [1,2] Even when such pairs are isolated (in radiolysis, several pairs may overlap forming a spur) their dynamics are fairly complex [3,4,5]. The simulation of these dynamics (given the multitude of other processes) is, therefore, difficult and cumbersome as compact analytical expressions do not exist and Monte Carlo or numerical methods should be used. [1] The situation becomes even more involved for multiple-pair spurs. A practical solution to this vexing problem are approximate methods based on "prescribing" the diffusion (see below). The method was originally suggested by Mozumder [6,7]; similar ideas and approaches form the basis of more complex models, e.g., the popular IRT model developed by Pimblott for simulation of water spurs. [8] Mozumder demonstrated [6] that for geminate partners migrating in the Coulomb field of each other the "prescribed" diffusion yields the same limiting survival probability as the exact theory. [3,5] When the pair is placed in the external electric field, the approximate solution given by the "prescribed diffusion" [7] is still reasonably close to the exact one. [3]

Does this recipe work for an arbitrary interaction potential? For example, for a mean field potential that is different from the Coulomb potential? Why does the "prescribed diffusion" approach gives the correct answer for the Coulomb potential? Is this approach correct? Does it yield reasonable asymptotic behavior? Given that the methods based on this "prescribed diffusion" approach are widely used to simulate complex dynamics in spurs, it is instructive to go back to the simplest case and find the answers to these questions.

The "prescribed diffusion" approach [6] seeks to find an approximate solution to equation [1-7]

$$\partial\rho/\partial t = D\nabla \cdot (\nabla\rho + \rho\nabla u), \tag{1}$$

where $\rho(\mathbf{r},\mathbf{r}_0;t)$ is the density function of a (single) geminate pair, $D$ is the mutual diffusion coefficient, $u(\mathbf{r}) = U(\mathbf{r})/k_B T$ is the reduced mean force potential, and $k_B T$ is the thermal energy. The point $\mathbf{r}_0$ is the starting point of the diffusion trajectory; one of the partners serves as the origin of the coordinate frame. Following Mozumder, [6,7] we will seek the solution of eq. (1) that has the form

$$\rho(\mathbf{r},\mathbf{r}_0;t) \approx \Omega(\mathbf{r}_0,t)\ P(\mathbf{r},\mathbf{r}_0;t), \tag{2}$$

where

$$P(\mathbf{r},\mathbf{r}_0;t) = (4\pi Dt)^{-3/2} \exp\left[-(\mathbf{r}-\mathbf{r}_0)^2/4Dt\right] \tag{3}$$

is the solution of equation (1) for $u = 0$ (i.e., free diffusion) that obeys the following normalization and boundary conditions

$$\oint_V d^3\mathbf{r}\ P(\mathbf{r},\mathbf{r}_0;t) = 1 \text{ and } P(\mathbf{r},\mathbf{r}_0;t\to 0) = \delta(\mathbf{r}-\mathbf{r}_0) \tag{4}$$

The survival probability $W(\mathbf{r}_0,t)$ of the geminate pair at time $t$ is therefore given by

$$W(\mathbf{r}_0,t) = \oint_V d^3\mathbf{r}\ \rho(\mathbf{r},\mathbf{r}_0;t) \approx \Omega(\mathbf{r}_0,t). \tag{5}$$

For Coulomb potential $u(r) = -r_c/r$, where $r_c$ is the Onsager radius of the potential. [3,4] In such a case, $\nabla^2 u = 0$ and equation (1) may be rewritten as

$$\frac{\partial\rho}{\partial t} = \frac{D}{r^2}\left(\frac{\partial}{\partial r}r^2\frac{\partial\rho}{\partial r} + r_c\frac{\partial\rho}{\partial r}\right). \tag{6}$$

Substituting eq. (2) into the latter formula and taking the integral over both parts of the resulting equation, [6] one obtains

$$\frac{\partial \Omega}{\partial t} = \Omega \times 4\pi D \int_0^\infty dr \left( r_c \frac{\partial P}{\partial r} \right) = \Omega \times f(t), \tag{7}$$

where

$$f(t) = (4\pi D)^{-1/2} r_c \ t^{-3/2} \exp\left[-r_0^2/4Dt\right]. \tag{8}$$

Using eq. (7) we find that

$$\Omega(r_0, t) = \exp\left(-\int_0^t f(t) \ dt\right). \tag{9}$$

Substituting $\xi = r_0/\sqrt{4Dt}$ into the latter formula, one obtains

$$\int_0^t f(t) \ dt = -\frac{r_c}{r_0} \int_\xi^\infty d\xi \ \exp(-\xi^2), \tag{10}$$

from which Mozumder obtained the following compact expression for the weighting function in eq. (2) [6]

$$\Omega(r_0, t) = \exp\left[-\frac{r_c}{r_0} erfc\left(\frac{r_0}{\sqrt{4\pi Dt}}\right)\right], \tag{11}$$

so that

$$\Omega_\infty(r_0) = \lim_{t \to \infty} \Omega_\infty(r_0, t) = \exp(-r_c/r_0) \tag{12}$$

For an arbitrary potential $u(r)$, substitution of eq. (2) into eq. (1) and averaging over the reaction volume gives

$$D^{-1}\frac{\partial \ln \Omega}{\partial t} = \langle \nabla \bullet P\ \nabla u \rangle = 4\pi \oint dr\ \frac{\partial}{\partial r}\left(P\ r^2\ \frac{\partial u}{\partial r}\right). \tag{13}$$

The integral on the right side is once more a complete differential (as in eq. (7)) and the right hand side of eq. (13) thereby equals $4\pi b P(a, r_0; t)$, where $a$ is the reaction radius and the parameter $b = a^2 (\partial u/\partial r)_a$ takes the role of the Onsager radius in eq. (8). Thus, we obtain

$$\Omega_\infty(r) = \exp(-b/r) \tag{14}$$

Formula (14) is certainly incorrect. It is easy to see that the limiting survival probability $\Psi(r) = \Omega_\infty(r)$ obeys the equation [1,5]

$$\nabla^2 \Psi = \nabla \Psi\ \nabla u, \tag{15}$$

from which

$$\frac{\partial}{\partial r}\left(r^2\ \frac{\partial \Psi}{\partial r}\right) = \frac{\partial u}{\partial r}\left(r^2\ \frac{\partial \Psi}{\partial r}\right) \tag{16}$$

and

$$\Omega_\infty(r) = r_c \int_a^r d\xi\ \xi^{-2}\ \exp[u(\xi)], \tag{17}$$

where

$$r_c^{-1} = \int_a^\infty d\xi\ \xi^{-2}\ \exp[u(\xi)] \tag{18}$$

defines (generalized) Onsager radius $r_c$ of potential $u(r)$. Equation (14) cannot be reduced to eq. (18) for any potential except for the Coulomb potential.

It is easy to see that the failure of the "prescribed diffusion" method is *conceptual* rather than mathematical, because eq. (2) does not have the commutation symmetry of the accurate solution. The solution of eq. (1) has the general property [5] that

$$\rho(\mathbf{r},\mathbf{r}_0;t) = \exp[u(\mathbf{r}) - u(\mathbf{r}_0)] \; \rho(\mathbf{r}_0,\mathbf{r};t) \tag{19}$$

Since for free diffusion $P(\mathbf{r},\mathbf{r}_0;t) = P(\mathbf{r}_0,\mathbf{r};t)$, combining eqs. (2) and (19) we obtain that

$$\Omega(\mathbf{r}_0;t) \; \exp[-u(\mathbf{r}_0)] = \Omega(\mathbf{r};t) \; \exp[-u(\mathbf{r})]. \tag{20}$$

As this equation holds for any $\mathbf{r}_0$,

$$\Omega(\mathbf{r};t) = \theta(t) \; \exp[u(\mathbf{r})] \tag{21}$$

where $\theta(t)$ is a function of time. Since $\Omega_\infty(r) \to 1$ for $u(r) \to 0$, $\theta(t \to \infty) = 1$ and

$$\Omega_\infty(r) = \exp[u(r)] \tag{22}$$

(compare with eq. (12)). The survival probabilities given by eqs. (17) and (22) should be equal. Equating these two expressions and taking the differential of both sides with respect to variable $r$, we obtain

$$\partial u / \partial r = r_c / r^2, \tag{23}$$

that is, $u(r) = -r_c/r$. In other words, the only potential for which the "prescribed diffusion" yields the correct estimate for the limiting survival probability is Coulomb potential. Furthermore, the correct answer obtained using this method for the Coulomb potential is purely accidental. The problem goes all the way back to eq. (2) of which eq. (22) is the immediate consequence. We conclude that eq. (2) does not generally hold: it is impossible to find a suitable function $\Omega(r_0,t)$ which approximates the exact solution, even at *infinitely long* delay time. This, in turn, means that the "prescription approach" does not generally work.

In conclusion, the prescribed diffusion approach does not work even for small deviations from the Coulomb potential. For the latter, the correct answer is obtained accidentally. Thus, extreme care should be exercised when "prescribed diffusion" approaches are used.